\documentclass[letterpaper]{article}
\usepackage{aaai2026}
\nocopyright
\usepackage{times}
\usepackage{helvet}
\usepackage{courier}
\usepackage[hyphens]{url}
\usepackage{graphicx}
\usepackage{natbib}
\usepackage{caption}
\usepackage{booktabs}
\usepackage{amsmath}
\usepackage{amssymb}
\graphicspath{{figures/}}
\newcommand{\tcite}{\ensuremath{\tau_{\mathrm{cite}}}}
\newcommand{\dollar}{\text{\textdollar}}
\usepackage{xcolor}

\title{Price Dislocations, News Citations, and Epistemic Leverage on Polymarket}
\author{Hazem Ibrahim\textsuperscript{\rm 1,*}, Yasir Zaki\textsuperscript{\rm 1}}
\affiliations{\textsuperscript{\rm 1}Computer Science, New York University Abu Dhabi, Abu Dhabi, UAE\\
\textsuperscript{*}Corresponding author: hazem.ibrahim@nyu.edu}

\begin{document}
\maketitle

\begin{abstract}
Prediction-market probabilities increasingly appear in news coverage, yet little is known about which market movements become news or how much trading money sits behind the numbers journalists quote. Unlike a public poll, a market price can be moved by anyone willing to trade, so the cost of manufacturing a number that later circulates as news weighs directly on the integrity of the information environment. Here, we answer these questions by linking 173.7 million signed Polymarket trades to news coverage from 2024--2025. From 6,990 English-language articles that mention prediction-market venues, an LLM-based, human-validated, sentence-level matcher extracts 1,582 sentences that quote market odds and attributes 918 of them to the specific market whose price they cite. We then detect 44,976 price dislocations, movements of at least five percentage points backed by concentrated one-sided trading, and analyze whether a market is cited more often afterward. In the days after a dislocation, a market's citation rate is about 33\% higher than its own matched baseline (market-averaged log citation-rate ratio $\tcite=0.283$, permutation $p=0.001$), and the same direction holds under binary and Poisson count versions of the outcome. The size of a dislocation, however, is not the strongest predictor of a media citation, as citations are associated far more strongly with how prominent the market already is (standardized $\beta=0.610$) than with how far the price moved ($\beta=0.159$). Finally, we combine the dollar flow that accompanies a given price change with observed citation rates into a descriptive metric we call \emph{epistemic leverage}, or the dollar amount needed to move a market five points and have the move cited. It stays near \dollar0.7--1.0 million across prominence quintiles, because markets that are cheaper to move are proportionally less likely to be cited. The threat model this implies centers not on the long tail of cheaply moved markets but on the small set of prominent markets that newsrooms already treat as informational infrastructure, and a seven-figure price of influence, prohibitive for most traders, sits well within the budgets of actors with a large stake in the number being quoted. We release aggregate event-study data and validation materials while withholding wallet-level and article-text records.
\end{abstract}

\section{Introduction}

Prediction markets have moved from specialist forecasting tools into the presentation layer of online news. A probability such as ``the market gives the candidate a 62\% chance'' can now appear beside a poll, a forecast model, or a conventional financial indicator. Polymarket and Dow Jones announced an exclusive arrangement covering \emph{The Wall Street Journal}, \emph{Barron's}, and MarketWatch, and rival venues have struck similar data partnerships with CNN, CNBC, and Fox \citep{polymarketdow2026,kalshicnn2025,kalshicnbc2025,kalshifox2026}. Prediction-market prices are therefore now becoming part of the standard toolkit news organizations use to describe uncertain futures.

A market probability, however, is produced differently from a poll estimate or a forecast-model output. It is a transaction price generated by participants with unequal capital, information, and incentives. Classical work emphasizes the ability of markets to pool information scattered across many traders \citep{wolfers2004,arrow2008}, while newer research documents concentrated ownership, disagreement between venues, and price impact on contemporary platforms \citep{clinton2025prediction,yangtsang2026,akey2026wins}. These facts do not make prediction-market prices uninformative; however, they do make the journalistic selection process important: when does a tradable number become a quoted fact?

This paper studies how a number produced by traders on one platform becomes content on another. A market probability is an aggregate signal that stands in for collective judgment, much as a trending topic, a review score, or a follower count does, but unlike those signals, it can in principle be moved by a single motivated participant. We ask three questions. First, are unusually large price movements, driven by concentrated trading, followed by increased citation of that market's odds? Second, when such movements are cited, is selection associated with the size of the move, the market's liquidity, or its prior prominence? Third, how does the estimated cost of moving a market compare with the observed chance that the movement is cited? To answer the third question, we define \emph{epistemic leverage}, a simple descriptive ratio that divides the estimated cost of a five-percentage-point movement by the observed seven-day citation rate. It plays the role that the price of purchased followers or coordinated upvotes plays in studies of platform manipulation, an empirical bound on how cheaply a signal of collective judgment can be manufactured. We interpret it descriptively, as the capital that the two measured quantities jointly imply, not as evidence that any observed movement is manipulated.

We answer these questions by linking 173.7 million signed Polymarket trades to a corpus of 6,990 articles. The core design is a pre-specified, self-controlled event study around 44,976 dislocations, price movements of at least five percentage points backed by concentrated one-sided trading, in which each market is compared with its own baseline windows. Our findings are further supplemented by a matched-market placebo, calendar-preserving permutations, a 2024 discovery/2025 confirmatory split, and a validated sentence-to-market matcher. Three findings organize the paper: 1) dislocations and media citations move together in time, but the strength of the link varies sharply by topic, and a shared news shock could produce the same pattern as media responding to prices, 2) move size has a statistically significant association with downstream citation, and 3) the cost of moving a market and the chance of being cited rise together with prominence. Obscure markers are cheap to move but rarely cited, whereas prominent markets are cited more often but are more expensive to move.

The setting is relevant to studies of the web because it is a cross-platform information-flow problem. A probability is produced by one web platform, picked up and reframed by newsrooms, and then distributed through search, social sharing, television graphics, and news homepages. The important question is therefore not only whether the market's final forecast turns out to be accurate. It is which of its intermediate prices become externally visible, under what market conditions, and how easily outsiders can audit them. By joining transaction data to textual citation, we measure this handoff rather than treating ``the media'' as an unobserved endpoint.

\paragraph{Contributions.} This study makes three contributions to web and social media research. (1) It provides the first large-scale linkage, to our knowledge, between transaction-level prediction-market movements and external news citation. (2) It develops and validates a reproducible measurement pipeline that maps odds-citing sentences to specific market questions. (3) It introduces epistemic leverage as a descriptive metric for reasoning about when platform-generated probabilities gain journalistic reach. The contribution is a measurement of a cross-system information pathway, not a causal estimate of manipulation or audience persuasion.

\section{Related Work and Conceptual Framework}

\paragraph{Prediction markets as information systems.} Prediction markets are commonly justified as mechanisms for aggregating dispersed beliefs into a price \citep{wolfers2004,arrow2008,hanson2003}. How to read those prices still depends on market design, participation, liquidity, and how prices map to beliefs \citep{manski2006,tetlock2008liquidity}. Studies of election markets often find them to be useful forecasts while also documenting disagreement and inefficiency \citep{berg2008,snowberg2013,clinton2025prediction}. Comparisons with polls and with structured human-judgment aggregation suggest that the forecasting advantage of markets is real but conditional, depending on who participates and what they know \citep{tetlock2015}. We therefore treat a market probability as a platform-produced signal whose informational value may be high without being mechanically objective.

\paragraph{Manipulation and price impact.} Classic theoretical and experimental work suggests that manipulation can be corrected by informed traders or can even subsidize information discovery \citep{hanson2006,rhode2006manipulating}. Market microstructure research treats price impact as a normal part of trading, not a sign that something is wrong. In the canonical models, a large order moves the price because other traders cannot rule out that the buyer knows something they do not \citep{kyle1985,glostenmilgrom1985}, and a long empirical literature measures how many dollars of trading it takes to move a price by a given amount \citep{hasbrouck1991,almgren2005direct,cont2014,kyleobizhaeva2016}. Our cost-to-move estimates apply this logic to prediction-market trades. Furthermore, recent work on large online venues reopens the empirical manipulation question. Deliberate experimental trades can move prices in ways that do not quickly revert, a small group of traders accounts for most holdings and profits, much of the activity comes from bots and professionals rather than retail bettors, and reconstructed Polymarket order flow shows that trades move prices by measurable amounts \citep{rasoolyrozzi2025,abid2026polymarket,yangtsang2026,akey2026wins}. We use this literature to motivate a price-impact proxy. However, we do not infer intent from a price movement and do not identify traders.

\paragraph{Manufactured signals on web platforms.} Social media research has long studied how signals of collective attention are produced and gamed. False content can travel farther and faster than true content \citep{vosoughi2018}, exposure to fabricated stories during the 2016 election was measurable at scale \citep{allcott2017}, and rumors can spread through social endorsement before verification \citep{sunstein2009}. The policy response to this literature centers on limiting coordinated manipulation of what appears organic \citep{lazer2018,aral2019}. A prediction-market price is a signal of the same family, produced by a smaller and better-capitalized crowd, with the distinctive property that moving it carries a posted dollar cost. That property is what makes leverage measurable here in a way it rarely is for followers, upvotes, or trends.

\paragraph{Probabilities as journalistic authority.} News organizations increasingly use data products to represent uncertain futures. Research on prospective data journalism shows that probabilistic displays can compress complex uncertainty into a small number of apparently precise indicators \citep{pentzold2020}. Agenda-setting research shows that the press confers salience by selection \citep{mccombs1972}, and work on exposure and persuasion documents how much reach and framing matter for what audiences take up \citep{zaller1992,bakshy2015,dellavigna2010}. In other words, marker odds cited by the media acquire a salience that unquoted ones lack, regardless of what readers ultimately believe. Criticism of prediction markets has recently focused on the possibility that a tradable number can be laundered into a neutral-looking forecast \citep{rohanifar2026,nechepurenko2026a,wein2026}. Those arguments identify a plausible pathway but do not measure how often it happens or which numbers get selected.

\paragraph{Unresolved prices as epistemic infrastructure.} Most empirical evaluation of prediction markets happens at the endpoint, once questions resolve, through calibration or profitability. Journalists do not encounter markets this way. They encounter a stream of unresolved prices, quote a small subset, and publish at a moment when no ground truth yet exists against which the number could be checked. The quoted price is therefore a snapshot the market itself may later revise. However, the journalist's article persists, and a move back in the odds does not necessarily trigger a correction in the copy. Endpoint metrics are blind to this gap, since a market can be well calibrated on average and still spend hours or days at a level produced by concentrated one-sided trading, and it is these intermediate levels, not the final resolution, that reach readers. The converse also holds, in that a cheaply moved market is a weak public threat if its price never leaves the platform. The newsroom partnerships motivating this study institutionalize exactly this sampling, as part of a broader platformization of knowledge in which external organizations consume platform-generated rankings, scores, and probabilities through APIs and feeds. That arrangement lowers the cost of reporting but can hide how a number was produced. The concern is not that the quoted prices are inaccurate; it is that a legible, continuously updated price can acquire institutional authority faster than its provenance travels. Our design therefore studies the selection step that sits between price production and public exposure, and epistemic leverage is an audit metric for that transfer.

\section{Research Questions and Scope of Claims}
\label{sec:questions}

\begin{table*}[t]
\centering
\small
\begin{tabular}{p{0.15\textwidth}p{0.18\textwidth}p{0.20\textwidth}p{0.38\textwidth}}
\toprule
Quantity & Unit and sample & Inferential status & Interpretation boundary \\
\midrule
Citation coupling $\tcite$ & Dislocation instance within a question-level market & Primary association estimand & How much higher (in log scale) a market's \emph{within-corpus} odds-citation rate is in an event window than in its own matched baseline; $\exp(\tcite)$ is the relative lift. A within-corpus ratio, not a share of all event coverage; shared news shocks remain possible. \\
Citation-selection coefficients & Detectable five-point dislocation event & Directional selection model & Conditional association of citation with move salience, flow-implied cost, and prior prominence; coefficients do not identify editorial motives. \\
Movement cost $C_i(5)$ & Outcome token with detectable positive impact & Microstructure component & Signed-flow amount associated with a five-point probability change under a linear realized-impact proxy; not historical order-book depth or manipulator loss. \\
Epistemic leverage $L_g$ & Category or prominence bin & Descriptive composite & Median movement cost divided by observed citation rate: a repeated-scenario accounting ratio, not a causal price of coverage. \\
\bottomrule
\end{tabular}
\caption{The paper's inferential hierarchy. Each quantity answers a different question and supports a
different level of claim.}
\label{tab:estimands}
\end{table*}

Our first research question (RQ) asks whether a price dislocation is followed by a change in citation relative to the same market's own baseline. The second RQ asks which attributes distinguish cited from uncited movements. Going in, we hypothesized that the size of the move (its ``salience'') would have a stronger effect on citation likelihood than the cost of producing the move (``liquidity''); prior prominence was also included as a competing explanation. The third RQ asks whether any markets are both cheap to move and likely to be cited. If such markets exist, they are a direct vulnerability. If none do, concern narrows to the few prominent markets that are expensive to move.

Each question is answered by a different quantity, and each quantity supports a different strength of claim. Table~\ref{tab:estimands} lists the unit, inferential status, and strongest permissible interpretation of each. For instance, even a strongly positive coupling estimate between dislocations and media citations can not definitively show that anyone manipulated a market, and the leverage ratio summarizes two observed quantities rather than naming a price at which news coverage could actually be bought.

\section{Data and Corpus Construction}
\label{sec:data}

\subsection{Market data}
Polymarket runs on the Polygon blockchain, and every trade is recorded publicly. We read trades directly from the blockchain's \texttt{OrderFilled} records rather than from the platform's public trade feed, because prior work shows the feed often gets the buy or sell direction wrong \citep{yangtsang2026}. Our analysis sample contains 173.7 million signed fills, trades whose buy or sell direction is known, across 224,637 outcome-token records. A catalog of 23,563 question-level markets has clear natural-language titles that we can match to news text. For robustness, we checked the panel of \citet{akey2026wins} against our own independent read of the Polygon blockchain and the buy or sell direction agreed on all 7,848 checked fills, with prices differing by an average of only $6.9\times10^{-6}$.

The inferential analysis covers the calendar years 2024 and 2025. We use 2024 to discover patterns and reserve 2025 to test whether they replicate. We exclude markets within 48 hours of resolution, when prices mechanically converge toward the final outcome. All price changes are expressed as changes in the implied probability of YES; a share price of \$0.62 implies a 62\% chance.

\subsection{News corpus}
We use GDELT, a global news database, to extract English-language articles that mention prediction-market venues during the analysis period \citep{leetaru2013}. After resolving links, removing exact duplicates, and downloading available article text, we are left with a corpus of 6,990 articles. Using regex patterns, we then flagged 1,826 candidate sentences that pair a venue or generic market reference with probability language, such as ``gives the candidate a 60\% chance.'' For each candidate sentence, we develop a retrieval system that ranks possible market questions by shared words (giving rare words more weight), event dates, named people and places, and the direction of the quoted probability. After human validation, we retained 1,582 genuine odds citation sentences and link 918 of them to one specific market, 905 of which are on Polymarket. The regulated venue Kalshi produced only 13 attributable odds sentences, so we disregard it in all analyses. Citations that cannot be tied to a specific market stay in corpus summaries but are excluded from market-level event studies.

We match sentences rather than whole articles because one article can mention several markets or compare multiple elections and economic indicators. We also split the matching into two steps. We first decide whether a sentence reports an odds value at all, and only then search the market catalog for the market it refers to. This split keeps a correct ``prediction markets say'' classification from being counted as a correct market match, and lets generic references (a sentence about ``betting markets'' in general) count as valid citations without matching to an incorrect market instance. Many outlets also republish the same wire-service or partner story under different links, and we remove such duplicates.

\subsection{Temporal alignment and topical exposure}
We timestamp market events from on-chain trades. From these trades we build two hourly series for each market. The first is the price, expressed as the implied YES probability; the second is buying pressure, the dollars spent buying minus the dollars spent selling. On the media side, we use the article's publication time when available and otherwise use the earliest reliable GDELT or page timestamp, and we conduct all comparisons at the daily level. The primary citation window analyzed is three days after the dislocation event, and we also test one- and seven-day windows for robustness (the full specification grid in Appendix~\ref{app:grid} reports every horizon).

A raw count of odds-citing articles mostly tracks how much news a topic generates, not how often journalists cite the market. Instead, we use a share. For each market $i$, we count every corpus article whose odds sentence the matcher shortlisted to $i$, whether or not the match was confirmed, and take the fraction of those articles that are confirmed citations of $i$. We call this fraction the \emph{within-corpus citation-intensity ratio}. This ratio captures how often journalists cite a market when they write about its topic at all. For instance, for the market ``Will Trump win the 2024 election?'', a raw count of odds-citing articles would surge in election week simply because election coverage surges. The ratio instead asks, among the shortlisted articles about that question in a given window, what fraction actually cite the market's odds, so a week of heavy election news with no extra odds-citing does not look like a citation spike. Because the corpus comes from a venue keyword search, the ratio is not the market's share of all news about the underlying event, and we never interpret it that way. Two caveats follow. First, the same retrieval step links both the candidate articles and the confirmed citations to markets, and the denominator therefore inherits matcher error (Section~\ref{sec:ethics}). Second, our primary results do not depend on this construction. For robustness, we also report a binary outcome (was the market cited at all), a Poisson count model with candidate exposure as a log offset (Section~\ref{sec:robustness}, Appendix~\ref{app:robust}), and a purged ratio whose denominator excludes the focal citation, keeping the numerator out of its own denominator.

\subsection{LLM-assisted annotation and human validation}
We used a language model (Claude Sonnet 5, Anthropic) to pre-label each candidate sentence on two questions: (1) Does the sentence cite prediction-market odds, and (2) is the retrieved market the correct match? We treat these model labels as measurement inputs, not ground truth. To validate them, two human annotators, neither shown the model's label, independently coded all 303 items of a stratified audit sample covering clear matches, weak matches, generic citations, and every candidate from the excluded venue. Appendix~\ref{app:validation} reports agreement between the annotators and precision under each one. Against the first annotator's labels, the model's odds-citation precision is 0.987 (Wilson 95\% interval $[0.967,0.995]$) and its market-attribution precision is 0.976 ($[0.945,0.990]$). The results are similar when considering the second annotator's labels (0.977 and 0.962). Because the audit samples items the machine flagged as positive, it estimates precision rather than end-to-end recall. We assess retrieval recall separately on a small hand-built pair set (Appendix~\ref{app:validation}).

\begin{table}[t]
\centering
\small
\begin{tabular}{p{0.45\columnwidth}r}
\toprule
Analysis object & Count \\
\midrule
Signed Polymarket fills & 173.7M \\
Outcome-token records & 224,637 \\
Question-level matching catalog & 23,563 \\
Fetched English news articles & 6,990 \\
Candidate odds sentences & 1,826 \\
Genuine odds citations & 1,582 \\
Citations linked to a specific market & 918 \\
Flow-concentrated dislocation events & 44,976 \\
Tokens in impact census & 11,898 \\
Events in citation-selection model & 16,236 \\
\bottomrule
\end{tabular}
\caption{Data products and principal analysis samples. Counts correspond to different stages and are
not intended to share a common denominator.}
\label{tab:samples}
\end{table}

\section{Measurement and Inference}
\label{sec:methods}

\subsection{Flow-concentrated dislocations}
A dislocation is defined as a price move of at least five percentage points driven by a burst of trading that pushes in the same direction, heavy net buying for a rise or heavy net selling for a fall. We call a burst ``heavy'' when its buying pressure ranks in the top 10\% of that market's own trailing 30-day distribution. We exclude events near resolution and merge adjacent qualifying intervals into a single episode. The primary detector produces 44,976 such dislocation events across 9,590 markets. Appendix~\ref{app:grid} lists the threshold, horizon, and denominator variants.

This detector flags unusual trading-and-price episodes; it does not identify manipulation. News can cause both the concentrated trading and the later citation, informed traders can act before journalists publish, and ordinary market making can move prices considerably within  thin markets. We use ``dislocation'' as a measurement label and reserve causal language for future work with designs that could identify intent or an outside shock to prices.

\subsection{Self-controlled citation coupling}
Our main analysis compares each market with itself. For market $i$, we count its odds-citing articles and its candidate articles inside the $H$-day windows after its dislocations ($A^{d}_i,T^{d}_i$), and inside baseline windows for the same market with no dislocation ($A^{0}_i,T^{0}_i$). Both kinds of window are whole calendar weeks, so a dislocation week is compared against the same market's ordinary weeks rather than against a different phase of the weekly news cycle, and baseline weeks are matched on topic and coarse coverage volume. Our estimate averages the difference between the two across markets,
\[
\tcite = \frac{\sum_i w_i\left[\log\frac{A^{d}_i+c}{T^{d}_i}-\log\frac{A^{0}_i+c}{T^{0}_i}\right]}{\sum_i w_i},
\qquad w_i=T^{d}_i+T^{0}_i,
\]
with pseudocount $c=0.5$, a small constant that keeps the logarithm defined when a count is zero. The weights $w_i$ give markets with more candidate articles more influence. $\tcite$ is a \emph{log rate ratio}. Taking $\exp(\tcite)$ tells us how many times higher a market's odds-citation rate is after a dislocation than in its own baseline weeks; it is not an absolute change in citations per article. A market enters our analytical sample only if it has both a dislocation-week and a baseline-week, and each market therefore serves as its own control.

Table~\ref{tab:flow} traces the sample at each step. The 44,976 detected dislocations span 9,590 markets, of which 3,496 also appear in the venue news corpus and form a 14,558-row market-week panel. Of those market-weeks, 8,669 carry a $\geq$5-point dislocation (the ``Events'' column of Table~\ref{tab:primarymain}) and 5,889 do not; 951 markets supply both states and therefore identify the within-market contrast. When a market-week has no candidate articles, we set its denominator to one so that it remains in the panel.

Because the same market appears in many windows, we cannot treat windows as independent observations. We instead test significance by permutation. We recompute $\tcite$ 1,000 times (with a fixed random seed) after shuffling which windows count as dislocation windows, only ever swapping labels within the same topic and calendar week. This preserves each market's structure and keeps every permuted dataset facing the same weekly news environment, since which topics were newsworthy in which weeks never changes, and averaging over markets keeps any single market from dominating. We also run a within-market shuffle and a comparison against matched markets that did not dislocate in the same topic-week. The design absorbs stable differences in how newsworthy a market is, but it cannot rule out a shared event that moves both the market and the press. $\tcite$ is therefore an association and not a causal effect.

\subsection{What predicts citation?}
For each detectable five-point dislocation, we model whether the market is cited within three days, using the logistic regression
\begin{align}
\operatorname{logit}\Pr(Y_{ie}=1)={}&\alpha_{c(i)}+\beta_s Z(|\Delta p_{ie}|) \notag\\
&+\beta_\ell Z(\log C_i(5))+\beta_v Z(\log V_{i,\mathrm{pre}}).
\end{align}
where $Y_{ie}$ indicates citation, $|\Delta p_{ie}|$ is the size of the price move (salience), $C_i(5)$ is the flow-implied cost of a five-point move (liquidity), $V_{i,\mathrm{pre}}$ is the market's prior dollar volume (prominence), and $Z(\cdot)$ puts all predictors on the same scale. Category fixed effects $\alpha_{c(i)}$ absorb broad topic differences, and we cluster standard errors by market \citep{cameronmiller2015}. This analysis includes 16,236 events.

\begin{figure*}[t]
\centering
\includegraphics[width=\textwidth]{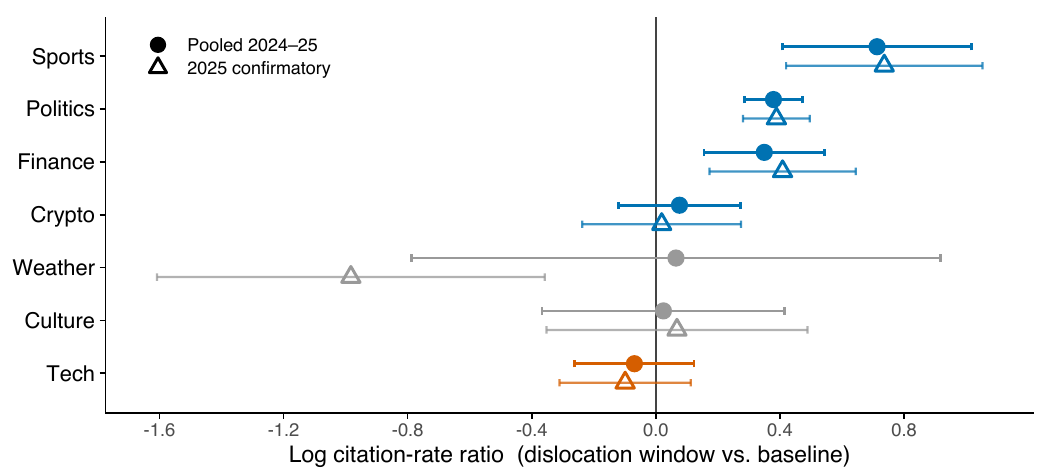}
\caption{Self-controlled citation coupling by category. Filled circles show the pooled 2024--2025 estimate;
open triangles show the 2025 confirmatory estimate; both carry 95\% intervals from the across-market
standard error. Each estimate is a market-averaged log citation-rate ratio between the three-day window
after a $\geq$5-point dislocation and the same market's matched baseline. Colored rows are rejected under BH-FDR in the full sample, where significance comes
from the calendar-preserving permutation test rather than from the plotted intervals; a colored row can
therefore have an interval that crosses zero. The formal 2024-to-2025 replication rule is reported in
Table~\ref{tab:primarymain}. The figure describes temporal association, not a causal effect of price on
coverage.}
\label{fig:coupling}
\end{figure*}

\subsection{Realized price impact and movement cost}
Full historical records of outstanding buy and sell orders at each price (the order book) is not publicly available for the period we study. We therefore estimate how much prices move per dollar traded from the trades that did execute. On Polymarket, each market's YES and NO positions trade as separate \emph{outcome tokens}, and we estimate impact per token. For a token $i$, we use non-dislocation intervals to fit
\[
\Delta p_{it}=\alpha_i+\lambda_i q_{it}+\varepsilon_{it},
\]
where $q_{it}$ is signed dollar flow and $\Delta p_{it}$ is the signed change in implied probability. We exclude a one-hour buffer around detected events. The fitted slope $\widehat{\lambda}_i$ measures how far the price has typically moved per dollar traded. A token counts as ``detectable'' when this slope is positive and statistically distinguishable from zero. For detectable tokens, we invert the slope to get $C_i(5)=0.05/\widehat{\lambda}_i$, the dollars that have historically accompanied a five-point move. We interpret $C_i(5)$ comparatively, as a measure of how much capital it takes to move one market relative to another.

\subsection{Epistemic leverage}
For each group of markets $g$, grouped by topic category or prominence bin, we define
\[
L_g=\frac{\operatorname{median}_{i\in g} C_i(5)}{\Pr(\text{event cited within 7 days}\mid g)}.
\]
We call $L_g$ an \emph{epistemic-leverage index}, the cost of moving prices divided by how often such moves get cited. The plotted intervals reflect uncertainty in the citation rate and, on the cost axis, a market-level bootstrap of the median; a joint bootstrap of the index itself widens the intervals without changing the pattern (Appendix~\ref{app:leverage}).

\begin{table*}[t]
\centering
\small
\begin{tabular}{lrrrrrrl}
\toprule
Category & Events & 2024 disc.\ $\tcite$ & 2025 conf.\ $\tcite$ & Conf.\ $p$ & Conf.\ BH $q$ & Pooled $\tcite$ & Repl. \\
\midrule
Sports   & 407   & $+0.910$ & $+0.736$ & $0.001$ & $0.001$ & $+0.713$ & yes \\
Politics & 4,457 & $+0.439$ & $+0.389$ & $0.001$ & $0.001$ & $+0.379$ & yes \\
Finance  & 822   & $+0.207$ & $+0.408$ & $0.001$ & $0.001$ & $+0.350$ & yes \\
Crypto   & 1,264 & $+0.276$ & $+0.019$ & $0.001$ & $0.001$ & $+0.076$ & yes \\
Culture  & 768   & $-0.900$ & $+0.068$ & $0.140$ & $0.163$ & $+0.024$ & no \\
Weather  & 83    & $+0.703$ & $-0.983$ & $1.000$ & $1.000$ & $+0.065$ & no \\
Tech     & 868   & $+0.242$ & $-0.099$ & $0.001$ & $0.001$ & $-0.069$ & no$^{\dagger}$ \\
\bottomrule
\end{tabular}
\caption{The seven-cell primary family. Estimates are market-averaged log citation-rate ratios
(dislocation window vs.\ the same market's matched baseline); $\exp(\tcite)$ is the multiplicative lift.
The confirmatory test is the 2025 estimate with its own permutation $p$ and BH $q$; the pooled
2024--2025 estimate is a more precise summary reported after replication, not the confirmatory test. Replication
requires concordant sign between the 2024 discovery and 2025 confirmatory sub-samples together with confirmatory
BH significance ($q<0.05$); the family and rule were fixed in advance (Appendix~\ref{app:grid}).
$^{\dagger}$Tech has a negative point estimate in the confirmatory and pooled samples but
a positive 2024 discovery estimate, so its sign is not concordant across the split. Four of seven
cells replicate (Sports, Politics, Finance, Crypto).}
\label{tab:primarymain}
\end{table*}

\subsection{Analysis plan and multiplicity}
All of our hypotheses and analyses follow a pre-specified plan. The specification grid, the primary family, the confirmatory rule, and the directional hypothesis were all fixed in written analysis-plan documents before estimation; Appendix~\ref{app:grid} records the plan in full and the documents are released verbatim with the replication materials.\footnote{Replication repository: \url{https://github.com/hazemibrahim97/epistemic-leverage-replication}.} The specification grid crosses venue, category, 5- and 10-point thresholds, 1-, 3-, and 7-day horizons, three outcome definitions, and raw versus purged news controls, for 756 reported cells. Because testing many cells produces some false positives by chance, we apply Benjamini--Hochberg false-discovery-rate control to the seven primary category cells as a separate family \citep{benjaminihochberg1995}. We kept the 2025 data out of the discovery set and use them as a confirmatory set to test whether the primary results would replicate. We count a result as replicated only if it keeps the same direction and stays significant after correction in the confirmatory set, not merely if confidence intervals overlap. We release the full grid allowing our results to be judged across the whole family rather than from a hand-picked window.

\section{Results}

\subsection{Dislocations are followed by elevated citation}
At the primary specification, a five-point move and a three-day window, dislocation events on Polymarket are more likely to be followed by a citation in the news. The estimate is $\tcite=0.283$ (permutation $p=0.001$), meaning a market's odds-citation rate is about $\exp(0.283)\approx1.33$ times, or a third higher than, its own matched baseline. Roughly 18\% of candidate articles in a covered window carry an odds citation within three days (2,313 of the 44,976 dislocations fall in a window with any such candidate article; Table~\ref{tab:flow}). The result is robust to how we construct the outcome. The contrast stays positive and significant when the outcome is a simple yes/no indicator of any citation (self-controlled permutation $p=0.001$) and when it is a Poisson count model with a log candidate-exposure offset (incidence-rate ratio $2.48$, $p=0.003$; negative binomial $2.55$, $p=0.003$). It is also stable across pseudocounts $c\in\{0.1,0.25,0.5,1.0\}$ ($+31\%$ to $+38\%$, all $p=0.001$; Appendix~\ref{app:robust}). Comparing against matched markets that did not dislocate also gives a positive difference (mean per-market difference in citation rate $0.434$, $p=0.001$, 242 matched pairs), and removing the cited articles themselves from the denominator barely changes the estimate ($0.276$ versus $0.274$). Together these checks make a mechanical artifact of the denominator unlikely.

This association differs significantly by topic (Fig.~\ref{fig:coupling}). Sports, Politics, Finance, and Crypto markets are positive and BH-significant in the full sample, keep the same sign in both the 2024 discovery and 2025 confirmatory sub-samples, and stay significant in the confirmatory sample. Tech, on the other hand, has a negative point estimate in the pooled and confirmatory samples while its 2024 estimate is positive; because the sign flips across the split, Tech does not count as a replication. Culture and Weather are not distinguishable from zero. In total, four of the seven primary categories (Sports, Politics, Finance, Crypto) replicate. A big move attracting coverage is therefore not a universal law, but rather, it depends on topic and period. One reason price dislocations in Tech markets may not be followed by press citations is that technology markets often track product announcements the press already covers directly, leaving reporters little reason to cite market odds, while political and sports markets offer a continuously updating probability for questions that get covered again and again.

\begin{figure}[t]
\centering
\includegraphics[width=\columnwidth]{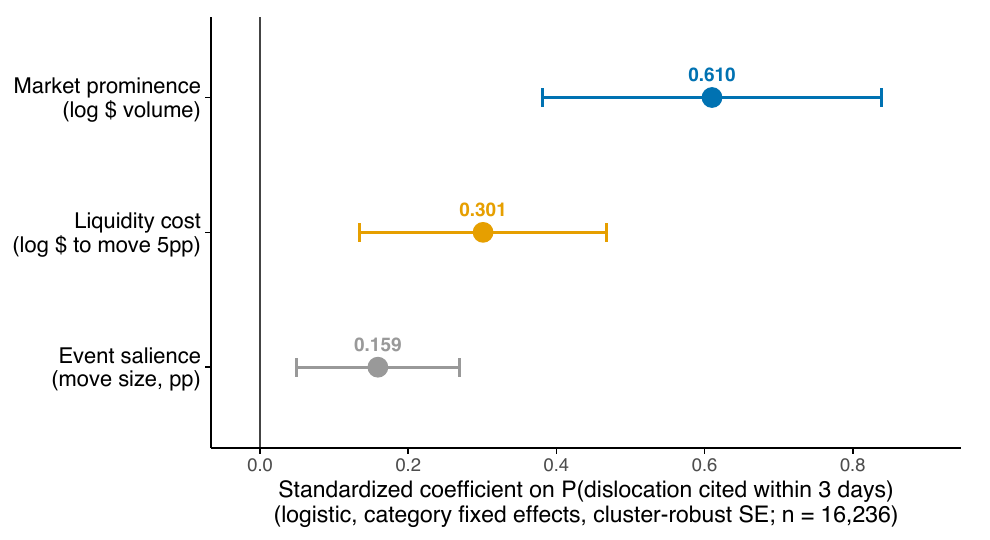}
\caption{Standardized coefficients from the logistic model of citation within three days of a
$\geq$5-point dislocation ($n=16{,}236$; category fixed effects; market-clustered standard errors).
We had predicted salience $>$ liquidity cost. Instead, prior market prominence is
the strongest correlate.}
\label{fig:ladder}
\end{figure}

\begin{figure*}[t]
\centering
\includegraphics[width=\textwidth]{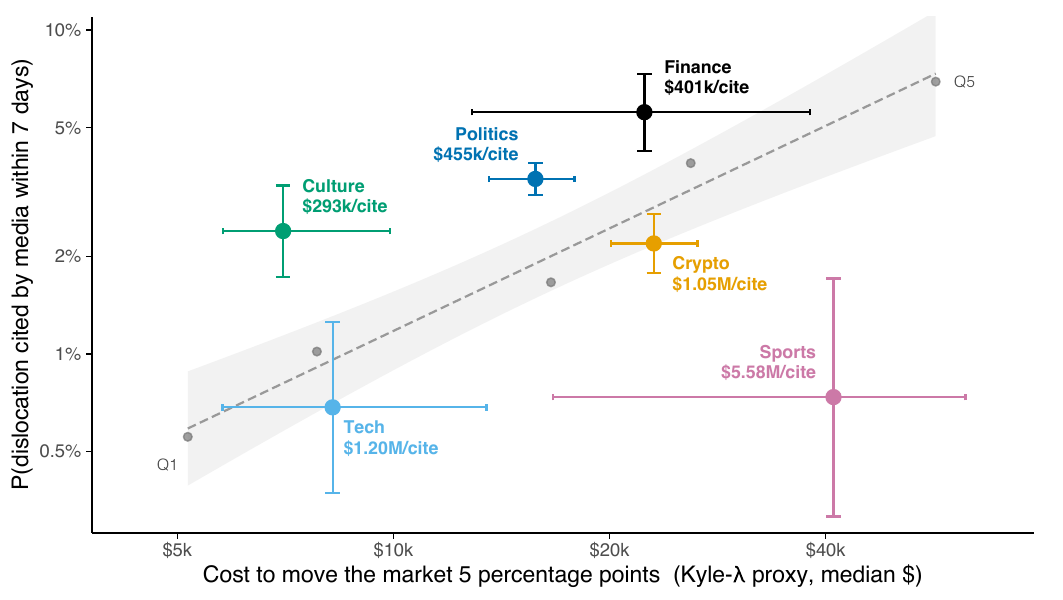}
\caption{The epistemic-leverage index on Polymarket. The horizontal axis is the median flow-implied
movement-cost proxy for a five-point move; the vertical axis is the observed seven-day citation rate.
Colored points represent categories, labeled with the index (movement-cost proxy divided by citation
rate); horizontal bars are market-level bootstrap 95\% CIs on the median cost and vertical bars are
Wilson 95\% CIs on the citation rate. Grey points are prominence quintiles (Q1 obscure to Q5 prominent),
with a dashed log-log fitted trend and its 95\% confidence band behind them. The index is
descriptively stable across prominence but varies widely by topic; it is an accounting ratio, not a
causal price of coverage.}
\label{fig:leverage}
\end{figure*}

\subsection{Journalistic selection is prominence-gated}
When a market dislocates, what decides whether the press picks it up? We had predicted that the size of the move would matter most. To test this, we fit the citation-selection model of Section~\ref{sec:methods}, which puts move size, movement cost, and prior market prominence on the same standardized scale and asks which best predicts citation within three days. The model does not support the ordering we hypothesized. Instead, market prominence has the largest standardized coefficient ($\beta=0.610$, SE $0.117$, $p<10^{-4}$), followed by liquidity cost ($\beta=0.301$, SE $0.085$, $p=4\times10^{-4}$) and move size ($\beta=0.159$, SE $0.056$, $p=0.005$; Fig.~\ref{fig:ladder}). All three are positive, but the prominence coefficient is nearly four times the move-size coefficient. A large move alone, therefore, is not enough to get media attention; the market must already be important enough to be on journalists' radar. We do not claim a causal ordering from these coefficients.

This pattern also helps explain the pooled event analysis. A dislocation can occur in thousands of low-volume markets without ever becoming news. Citations, instead, concentrate among markets that are already visible, liquid, and topically consequential. The pattern is compatible with editorial judgment, search and ranking effects, venue promotion, and newsroom partnerships.

\subsection{The leverage index is approximately stable across prominence}
How many dollars does it take to move a market \emph{and} have the move show up in the news? This question has two halves, what a price move costs and how often a move of that size reaches the press. We measure the first half with the realized price-impact model, the second with the seven-day citation rate among dislocation events, and we divide the two into the leverage index.

For the cost half, we fit the price-impact model to every eligible token and inverted the fitted slope into a movement-cost proxy. Among the 11,898 tokens in the impact census, 43\% have a detectable positive impact coefficient. Among these detectable tokens, the median flow-implied cost of a five-point move is \dollar21,773, and \dollar43,547 for ten points. Category medians range from \dollar6,800 in Culture and \dollar8,144 in Tech to \dollar110,183 in Sports (Appendix~\ref{app:impact}). These values summarize how much money has historically accompanied a given price change outside detected episodes. A non-detectable coefficient can mean noise, too few trades, nonlinear impact, or a deep market. It does not mean the market costs nothing, or infinitely much, to move. All leverage comparisons therefore use the same detectable population and compare within it.

For the citation half, only about 1.4\% of dislocation events are cited within seven days, against a median five-point movement cost of \dollar16,485 among these tokens. Dividing the cost by the citation rate gives the leverage index. At the category level it ranges from \dollar293,000 in Culture and \dollar455,000 in Politics to \dollar5.58 million in Sports. This large spread across topics reflects how differently newsrooms demand market-generated probabilities, not just how deep the markets are.

One concerning scenario is an obscure market that offers cheap influence. Thin markets cost the least to move, so if the press cited their moves at anything close to the rate of prominent markets, a few thousand dollars would buy a citable statistic, a manufactured probability that then circulates in the news with the outlet's credibility attached rather than the buyer's. To test whether such a pocket exists, we recomputed the index across prominence quintiles (Fig.~\ref{fig:leverage}, gray line; Table~\ref{tab:quintile}). We find that, from the least to the most prominent quintile, the median movement cost rises about 11-fold, from \dollar5,171 to \dollar56,927, and the citation rate rises almost in step, about 12.5-fold, from 0.55\% to 6.93\%. Because the two rise together, their ratio stays within a \dollar0.7--1.0 million band. Therefore, a market that is cheap to move is also unlikely to be cited, and by almost the same factor. A market-level joint bootstrap that resamples both components formalizes this phenomenon. The cross-quintile spread of the index is 1.49$\times$ (95\% CI [1.32, 4.37]), still an order of magnitude below the 19$\times$ spread across categories (Appendix~\ref{app:leverage}). The index is therefore not exactly invariant. Nonetheless, we find no group of markets that are both cheap to move and widely cited.

\section{Robustness and Alternative Explanations}
\label{sec:robustness}

The pattern we measure can arise through several pathways. A public event can move both prices and coverage; informed traders can act before journalists publish; reporters can react to the market; or a platform can promote a moving contract and thereby raise both trading and visibility. No observational comparison of web traces can fully separate these mechanisms. Our robustness analyses therefore focus on a narrower set of mechanisms. Specifically, they test whether the measured coupling is an artifact of baseline choice, denominator construction, matcher error, multiple testing, or a single calendar period.

Appendix~\ref{app:robust} reports the full set of checks; here we summarize what each one tests and finds. Same-market baseline weeks absorb stable newsworthiness, and 1,000 calendar-preserving topic-week permutations put the pooled estimate far outside its reference distribution ($\tcite=0.283$, $p=0.001$). Matched markets in the same topic and week that did not dislocate are cited less than the dislocated markets (contrast $0.434$, $p=0.001$, 242 pairs). Removing the focal odds-citing article from the exposure denominator leaves the estimate essentially unchanged ($0.276$ versus $0.274$), and counting only citations on strictly later calendar days than the move keeps the coupling ($\tcite=0.239$, $p=0.001$; 74\% of first citations follow the dislocation). The 2024 discovery to 2025 confirmatory split replicates four of the seven primary category cells, and the blinded 303-item double-coded audit bounds matcher precision at $0.987$ and $0.976$. Table~\ref{tab:robustness} in the appendix pairs each threat with its diagnostic and with what remains unaddressed. None of these checks identifies causation. A shared event that moves one market more strongly than its matched controls remains possible, and we therefore avoid verbs such as ``drives,'' ``causes,'' or ``induces'' when describing $\tcite$.

\section{Discussion}

This study began by asking whether a thin, tradable probability can be moved and then laundered through news coverage as objective fact. The empirical answer is narrower than that rhetoric suggests. Price dislocations and citation are coupled in several high-attention domains, but the size of the move is the weakest of the three predictors while the market's prior prominence dominates. The markets most likely to become news are therefore exactly the markets that cost the most to move.

This changes where concern about manipulation should focus. The worry that motivated the study is the long tail, thousands of thin contracts whose prices cost almost nothing to move. Our results suggest the tail is not where the exposure is, because a manipulated price only matters once someone repeats it, and the press rarely looks at obscure markets. The roughly constant leverage index across prominence quintiles shows that these two costs offset each other, and the far wider spread across topics shows that newsroom demand for probabilistic updates matters at least as much as market depth. The exposed surface is the small set of prominent markets in topics journalists already cover. One caution applies when reading the cost figures this way. The proxy measures the money that accompanied price changes, not a manipulator's profit, so it identifies which markets are cheap to move, not whether moving them pays.

However, the stability of the index should not be read as reassurance. While a seven-figure price is prohibitive for most, it is nonetheless small for anyone with a serious stake in the quoted number, and campaigns, industry groups, and governments routinely spend more than that on channels which carry less credibility than a probability reported as fact by a national outlet. What the index prices is therefore not safety but the identity of the plausible buyer. A near-zero cost would open the channel to anyone, while a seven-figure cost reserves it for well-resourced and motivated actors. The figure is also a gross cost. A trader who moves a price ends up holding a position, not spending the money, and can later sell it back, so the true cost of the influence is only a fraction of the flow we measure.

For journalists, the practical implication concerns attribution. A market price is not a neutral instrument reading. It is produced by whoever chose to trade, and our movement-cost estimates show that in many markets the money required to move it is modest. Quoting the odds is therefore closer to quoting a source than to citing a measurement, and it deserves the same provenance. A minimal practice would report the venue, the contract wording, the timestamp, recent volume, and whether the move sat in an extreme flow percentile, the analogue of a poll's sample size and field dates. Whether displaying that provenance changes what readers believe is a question for future experiments rather than for our observational data. Selection on prominence also carries a feedback risk. Reporters cite visible contracts, and coverage makes contracts more visible. We cannot estimate this feedback loop here, but the strong prominence coefficient shows where it would operate.

For platforms and regulators, the study separates public auditability from regulation. We could reconstruct signed flow, timestamps, and price formation on Polymarket only because settlement is public, and that openness is what makes outside audits like this one possible at all. A venue does not need to expose its traders to offer the necessary data to make such inferences possible. Releasing historical prices, aggregate signed flow, volume, resolution-rule changes, and stable contract identifiers would let outsiders estimate impact and flag unusual flow without any trader-level records, and news partnerships could require that layer by contract.

For studies of the web, our contribution is treating citation between platforms as something to measure directly. We link a specific market question to a specific news sentence rather than using news volume as a background control, and the same recipe applies to other platform numbers that get quoted as fact, from cryptocurrency prices to trend scores and polling aggregates. The validation lesson is to split the measurement. Detecting an odds citation and attributing it to the right market fail in different ways, so we validate them separately and let their error rates bound our claims.

What would it take to show influence rather than association? A stronger design would move the displayed probability while holding the underlying news constant, through market-maker outages, display changes, contract migrations, or staggered newsroom integrations, or would randomize what editors see. None of these levers are available to outside researchers working from public traces. They require either the platform's cooperation or a newsroom partner, and the natural experiments that could substitute, such as the newsroom data integrations, mostly launched after our observation window closed. Measuring the coupling observationally is the step that such designs would build on, and the integration dates are a ready-made quasi-experiment once enough post-launch data accumulates. Whether citation then changes what audiences believe is a further step still, and would need exposure experiments. We keep these stages separate so that a robust price-to-citation association is not understood as a claim about public belief.

\section{Limitations, Ethics, and Responsible Release}
\label{sec:ethics}

The study has five central limitations. First, the coupling design is observational; it cannot remove shared news shocks or establish whether prices moved before journalists knew the same facts. Second, realized price impact is not order-book depth and is detectable for only 43\% of eligible tokens. Third, the corpus is English-language news found through a \emph{venue keyword} search on GDELT and available web pages, and paywalls, dead links, syndication, and outlet indexing create coverage bias. The topical denominator lives inside this venue corpus and comes from the matcher's retrieval shortlist, and $\tcite$ therefore measures citation intensity among the corpus's odds-candidate mentions of a market, not the market's share of all news about the underlying event. Fourth, the annotation audit covers all 303 sampled candidates and was independently coded by a second annotator, but it estimates precision more directly than recall because it samples machine-positive candidates, a constraint inherent to auditing a running matcher. Fifth, the 2024/2025 split is a temporal replication inside one rapidly changing platform ecosystem, not a replication across countries or media systems.

These limitations are consequential with respect to different claims. Shared shocks limit the causal reading of the coupling. The within-corpus denominator limits what $\tcite$ measures, though the prominence model and the leverage composite use per-event citation indicators rather than the topical rate and are unaffected by it. Impact detectability limits which tokens the cost results cover, and corpus and matcher coverage limit the media-selection estimates. None of these is fixed by simply collecting more dislocation events. The paper therefore reports sample counts at each stage and makes no single ``complete data'' claim across the pipeline.

All inputs are public secondary data, and the study involves no intervention, no deanonymization, and no contact with human subjects. Polymarket transactions come from the public Polygon blockchain and the reused CC-BY signed-fill panel of \citet{akey2026wins}; market metadata comes from Polymarket's public Gamma endpoints; and the news corpus is built from GDELT and publicly reachable article pages \citep{leetaru2013}. We release \emph{only aggregated or time-coarsened outputs}: the 756-cell result grid, category- and quintile-level tables, aggregate dislocation--citation event records, annotation instructions, and gold-set judgments. We redistribute no raw article text and no wallet-level records. Because blockchain addresses are pseudonymous and potentially re-identifiable, we deliberately do not publish wallet-level labels, market-specific ``cheapest target'' tables, or any records that join episodes to individual traders. These restrictions reduce individual-event reproducibility but mitigate defamation, privacy, and dual-use risks. The work is adjacent to gambling harm and election integrity \citep{packin2026}; accordingly, we avoid identifying alleged manipulators and avoid operational instructions.

\bibliography{references}

\clearpage
\appendix

\section{Analysis grid, pre-specified plan, and primary family}
\label{app:grid}
\paragraph{The analysis plan, its lineage, and what was fixed in advance.} The confirmatory plan was fixed in written specification documents before the coupling estimates were computed, and those documents are released verbatim with the replication materials. It fixed the 2024--2025 analysis window, the 2024 discovery / 2025 confirmatory split, the five- and ten-point dislocation thresholds, the one-, three-, and seven-day citation horizons, the three outcome forms, the focal-article purge, the seven-cell primary family, and the directional salience-versus-liquidity hypothesis. We also distinguish pre-specified tests from result-driven interpretation (Table~\ref{tab:preregmap}). The prominence coefficient was a specified competing predictor, but the emphasis on a ``prominence gate'' follows from its realized magnitude. The description ``approximately stable across prominence'' summarizes the observed quintile pattern and is not a pre-specified equivalence or scale-invariance claim. Category-level discussion beyond the seven-cell family is descriptive unless explicitly identified as a corrected primary result. This record preserves the value of the plan without pretending that all language in the final paper was forecast in advance.

The pre-specified grid crosses venue, category, dislocation threshold $\{5,10\}$ percentage points, citation horizon $\{1,3,7\}$ days, outcome $\{$citation count, citations per topical article, binary cited$\}$, and news control $\{$raw, focal-citation purged$\}$. The implementation writes 756 cells to a machine-readable table. The seven Polymarket, five-point, three-day, citations-per-topical-article category cells form the primary family; BH-FDR is applied within that family. Table~\ref{tab:primarymain} in the main paper reports those cells. All remaining cells are robustness or descriptive analyses rather than silently selected alternatives.

\begin{table*}[t]
\centering
\small
\begin{tabular}{p{0.62\columnwidth}r}
\toprule
Stage (unit) & Count \\
\midrule
Detected flow-concentrated dislocations (event) & 44,976 \\
\quad distinct markets they span (market) & 9,590 \\
\quad markets also present in the venue news corpus (market) & 3,496 \\
Coupling analysis panel (market-week) & 14,558 \\
\quad $\geq$5pp dislocation-weeks $=$ ``Events'' in Table~\ref{tab:primarymain} & 8,669 \\
\quad baseline (non-dislocation) weeks & 5,889 \\
\quad markets with both states (identify the estimator) & 951 \\
Descriptive: dislocations with $\geq$1 within-corpus candidate article & 2,313 \\
\bottomrule
\end{tabular}
\caption{Sample flow for the self-controlled coupling estimate. The analysis unit is the market-week, not
the raw dislocation: the ``Events'' column of Table~\ref{tab:primarymain} counts $\geq$5-point
dislocation-weeks (summing to 8,669 across the seven categories), while the within-market contrast is
identified by the 951 markets that contribute both a dislocation-week and a baseline-week. The 2,313 figure
is a descriptive coverage count (dislocations whose window contains any candidate article), not the rate
estimand's denominator, which is floored at one so every market-week is retained.}
\label{tab:flow}
\end{table*}

\begin{table*}[t]
\centering
\small
\begin{tabular}{p{0.32\textwidth}p{0.25\textwidth}p{0.32\textwidth}}
\toprule
Design element & Plan status & Final-paper treatment \\
\midrule
2024 discovery / 2025 confirmatory & Pre-specified & Confirmatory set used only after primary discovery specification fixed \\
5/10pp thresholds; 1/3/7d horizons & Pre-specified grid & Five-point, three-day cell is primary; all variants released \\
Three outcome forms and focal-article purge & Pre-specified grid & Rate is headline; count, binary, and purged variants are robustness \\
Seven category cells with BH-FDR & Pre-specified family & Reported in full in Table~\ref{tab:primarymain} \\
Salience $>$ liquidity directional hypothesis & Pre-specified & Not supported; prominence, a specified competitor, is largest \\
``Prominence gate'' emphasis & Result-driven interpretation & Labeled as interpretation, not a forecast claim \\
Approximate scale invariance & Result-driven summary of pre-specified composite & Descriptive; no equivalence test claimed \\
\bottomrule
\end{tabular}
\caption{Plan-to-paper map: which design elements were fixed in the analysis plan and how the
final paper treats each.}
\label{tab:preregmap}
\end{table*}

The discovery/confirmatory split is calendar-based rather than random. This is a demanding test with respect to platform growth and newsroom adoption because the 2025 environment differs from 2024, but it does not guarantee transport to 2026 or later. ``Replication'' in the paper means concordant direction and corrected significance under the registered rule, not independent reproduction by another team.

\paragraph{Pipeline chronology and what the confirmatory split does and does not test.} Honesty about the split requires separating the confirmatory \emph{statistical} plan from the \emph{measurement} pipeline. The order was: (i) collect the full 2024--2025 market and news data; (ii) develop the measurement pipeline (dislocation detector, sentence matcher and its retrieval features, topical-shortlist construction, category assignment, and impact-model eligibility) on that full corpus; (iii) fix in writing the confirmatory grid, primary family, thresholds, horizons, the discovery/confirmatory split, and the salience-versus-liquidity hypothesis; and (iv) compute the coupling grid, first on 2024 for discovery and then on 2025. The dislocation thresholds and collapsing rules, impact-model fill bounds, baseline-window matching, and near-resolution exclusions were fixed by the registration rather than tuned to either year. The matcher and topical-matching procedures, however, were built on the full corpus, and 2025 articles were visible during measurement development. The 2025 year is therefore a genuine out-of-sample test for the coupling \emph{estimation}, which ran only after the split was fixed, but it is a temporal robustness sample rather than a fully untouched confirmatory set for the measurement instruments themselves. Freezing the pipeline on 2024 alone and re-running it blind on 2025 is a natural design for future work.

\section{Corpus construction and matcher validation}
\label{app:validation}

\subsection{Retrieval stages}
The media pipeline has four stages: (1) venue-keyword discovery in GDELT; (2) URL resolution, deduplication, and article fetching; (3) sentence extraction using venue, probability, and odds language; and (4) sentence-to-market retrieval over the 23,563-question catalog. Retrieval is anchored to the candidate sentence rather than the full article because many articles mention several unrelated events. Ranking uses overlap between the sentence and the market question, rarity weighting of shared words, named entities, date compatibility, direction words, and probability values. Markets of the form ``will person say word X'' are excluded from retrieval because their wording overlaps with too many unrelated sentences and creates systematic false matches.

\subsection{Audit design}
The validation frame contains 303 machine-positive items from four strata: clear specific-market citations, weak or low-confidence specific-market citations, generic citations with no specific market, and all 13 candidates from the excluded venue (Kalshi). Each annotator saw the article sentence and the retrieved market but not the model's label, and \emph{independently} coded every item on both questions (is it an odds citation, and is the market correct) before any reconciliation; this is a two-coder design, not a review of one coder's labels. We use Wilson intervals because some strata are small. Because the frame samples machine-positive candidates, it bounds precision, not recall. The topical denominator (Section~\ref{sec:data}) inherits the same retrieval linkage, and future work should validate it separately.

\paragraph{Inter-annotator agreement.} The two coders agree on 96.4\% of items for the odds-citation judgment (11 disagreements) and 93.7\% for market attribution (19 disagreements). Because the gold set is deliberately drawn from machine positives, roughly 99\% of items are labeled ``yes'' by both coders and the two never jointly label an item ``no.'' Under such lopsided prevalence, Cohen's $\kappa$ breaks down; expected agreement ($0.964$) essentially equals observed agreement, giving $\kappa\approx0$. This is the well-known high-agreement/low-$\kappa$ paradox, not evidence of chance labeling \citep{feinstein1990,gwet2008}. We therefore report agreement measures that are robust to prevalence. Gwet's AC1 is $0.962$ ($[0.937,0.983]$) for odds citation and $0.933$ ($[0.899,0.962]$) for attribution, and the prevalence-and-bias-adjusted $\kappa$ (PABAK) is $0.927$ and $0.875$; all indicate near-perfect agreement. Table~\ref{tab:matcher} reports precision against the first annotator. Using the second annotator's independent labels gives $0.977$ odds-citation and $0.962$ attribution precision, and the validation therefore does not depend on which coder is treated as the reference.

\begin{table*}[t]
\centering
\small
\begin{tabular}{lrcc}
\toprule
Stratum & $n$ & Odds-citation precision & Attribution precision \\
\midrule
Clear citation    & 110 & $0.982\ [0.94,1.00]$ & $0.981\ [0.94,1.00]$ \\
Weak/low confidence & 90 & $0.989\ [0.94,1.00]$ & $0.966\ [0.91,0.99]$ \\
Generic citation  & 90 & $0.989\ [0.94,1.00]$ & N/A \\
Excluded venue    & 13 & $1.000\ [0.77,1.00]$ & $1.000\ [0.77,1.00]$ \\
\midrule
Stratum-weighted overall & 303 & $\mathbf{0.987\ [0.97,0.99]}$ & $\mathbf{0.976\ [0.95,0.99]}$ \\
\bottomrule
\end{tabular}
\caption{Blinded audit of machine-positive candidate citations, all 303 items independently double-coded by
two annotators; precision is shown against the first annotator (inter-annotator agreement in text). Overall
intervals are pooled Wilson intervals (odds citation $299/303$; attribution $205/210$). Attribution
precision is undefined for generic citations.}
\label{tab:matcher}
\end{table*}

The matcher's nine errors against the first annotator split into four sentences the human judged not to assert market odds (a stray probability not sourced to a market, or an aggregator restatement) and five whose number referred to a different market than the retrieved question, several in weak-stratum cases where the true market was missing from the shortlist. If false positives are spread roughly evenly over time, they wash out in self-controlled comparisons. Errors that cluster around events remain possible; this motivates the matched-control and confidence-stratum sensitivities.

The separate retrieval test contains 21 hand-identified article--market pairs. The final shortlist placed all hard cases and 94\% of easy cases within the candidate set. Because this test is small and was used during retrieval development, it is a diagnostic, not an unbiased estimate of production recall.

\section{Price-impact construction}
\label{app:impact}
We estimate token-level impact only on non-dislocation observations, with a one-hour buffer around detected episodes. Tokens must have between 1,000 and 50,000 fills, which avoids very sparse fits and keeps a few extremely active tokens from dominating the computation. Of 11,898 fitted tokens, 43\% have a positive, statistically resolvable coefficient. All reported medians condition on this detectable subset.

\begin{table*}[t]
\centering
\small
\begin{tabular}{lrrr}
\toprule
Category & Tokens & Median $\Delta p$ / \dollar10k & Median \dollar / 5pp \\
\midrule
Culture  & 668  & $0.116$ & \dollar6,800 \\
Tech     & 418  & $0.089$ & \dollar8,144 \\
Politics & 1,763 & $0.040$ & \dollar20,750 \\
Crypto   & 1,257 & $0.037$ & \dollar24,281 \\
Finance  & 206  & $0.033$ & \dollar27,821 \\
Sports   & 766  & $0.010$ & \dollar110,183 \\
\bottomrule
\end{tabular}
\caption{Realized price-impact census for detectable tokens. The pooled median is \dollar21,773 for a
five-point movement and \dollar43,547 for ten points.}
\label{tab:lambda}
\end{table*}

Three qualifications matter. First, the linear fit is a local summary; impact can be nonlinear, state-dependent, and asymmetric near the probability boundaries. Second, executed flow is not the same as the cost of eating through a live order book, because new limit orders can arrive while a trade sequence unfolds. Third, detectability is a selected sample. We therefore use impact mainly for rank and ratio comparisons within a common sample, not for market-specific operational estimates.

\section{Epistemic-leverage decomposition}
\label{app:leverage}
Table~\ref{tab:leverage} splits the ratio into its two measured components. The event-conditioned movement-cost median is lower than the full-census median because observed dislocations happen disproportionately in tokens that are easy to move. Weather has no citation in the relevant window, and its ratio is therefore undefined rather than infinite.

\begin{table*}[t]
\centering
\small
\begin{tabular}{lrrrr}
\toprule
Category & Events & \dollar/5pp & Cited within 7d & $L$ (\dollar/citation) \\
\midrule
Culture  & 4,964  & \dollar7,020  & $2.39\%$ & \dollar293k \\
Finance  & 2,732  & \dollar22,363 & $5.57\%$ & \dollar401k \\
Politics & 20,054 & \dollar15,766 & $3.47\%$ & \dollar455k \\
Crypto   & 9,042  & \dollar23,048 & $2.19\%$ & \dollar1.05M \\
Tech     & 3,725  & \dollar8,228  & $0.68\%$ & \dollar1.20M \\
Sports   & 3,881  & \dollar41,014 & $0.74\%$ & \dollar5.58M \\
Weather  & 578    & \dollar16,488 & $0.00\%$ & undefined \\
\bottomrule
\end{tabular}
\caption{Category decomposition of epistemic leverage for detectable Polymarket dislocation events.}
\label{tab:leverage}
\end{table*}

Table~\ref{tab:quintile} gives the same decomposition across prominence quintiles (based on prior fills) of the detectable dislocation events. Movement cost and citation rate rise together, and the index stays within a narrow band. The reported intervals are Wilson intervals on the citation rate. A market-level joint bootstrap (2,000 draws; markets resampled with replacement; cost median and citation rate recomputed jointly per draw; quintile bins held fixed) gives quintile leverage intervals of \dollar933k [\dollar535k, \dollar2.38M] for Q1, \dollar770k [\dollar448k, \dollar1.63M] for Q2, \dollar997k [\dollar701k, \dollar1.63M] for Q3, \dollar669k [\dollar428k, \dollar1.18M] for Q4, and \dollar822k [\dollar588k, \dollar1.21M] for Q5, with a cross-quintile spread of 1.49$\times$ (95\% CI [1.32, 4.37]). A draw in which a resampled group has zero citations makes the index infinite; such draws occur for Sports (101 of 2,000) and Weather (all draws) and are excluded from the percentile intervals.

\begin{table*}[t]
\centering
\small
\begin{tabular}{llrrlr}
\toprule
Quintile & Fills range & $n$ & \dollar/5pp & Cited within 7d [95\% CI] & $L$ (\dollar/citation) \\
\midrule
Q1 & 1{,}000--1{,}981   & 3,247 & \dollar5,171  & $0.55\%\ [0.35,0.87]$ & \dollar933k \\
Q2 & 1{,}981--3{,}292   & 3,247 & \dollar7,824  & $1.02\%\ [0.72,1.42]$ & \dollar770k \\
Q3 & 3{,}292--5{,}802   & 3,247 & \dollar16,573 & $1.66\%\ [1.28,2.16]$ & \dollar997k \\
Q4 & 5{,}802--11{,}808  & 3,247 & \dollar25,947 & $3.88\%\ [3.27,4.60]$ & \dollar669k \\
Q5 & 11{,}808--49{,}967 & 3,248 & \dollar56,927 & $6.93\%\ [6.10,7.85]$ & \dollar822k \\
\bottomrule
\end{tabular}
\caption{Epistemic-leverage index by market-prominence quintile (equal-count bins of detectable Polymarket
dislocation events, split on prior token fills). Cost rises $11\times$ and citation rate $12.5\times$ across
the range, so the index varies only $\sim$$1.5\times$.}
\label{tab:quintile}
\end{table*}

The category spread in $L$ is roughly 19-fold, much wider than the 1.5-fold spread across prominence quintiles. This contrast is why we describe topic-level newsroom demand as the larger source of variation. The ratio should not be averaged across categories without weighting, because category composition changes over time. Each quintile summary rests on five aggregated points, and we therefore treat the cross-quintile stability as a descriptive pattern; alternative impact specifications (probability-bin controls, separate up/down estimates) are a natural extension for future work.

\section{Additional robustness and scope checks}
\label{app:robust}

\begin{table*}[t]
\centering
\small
\begin{tabular}{p{0.20\textwidth}p{0.30\textwidth}p{0.39\textwidth}}
\toprule
Threat & Diagnostic & Result and remaining interpretation \\
\midrule
Stable market newsworthiness & Same-market matched offset windows & Absorbs time-invariant prominence and topic differences; cannot absorb time-varying news. \\
General topic-week shock & 1,000 calendar-preserving topic-week permutations & Pooled $\tcite=0.283$, $p=0.001$; timing is unusual relative to the preserved calendar structure. \\
Threshold-independent news & Matched non-dislocated markets in the same topic-week & Positive contrast $0.434$ ($p=0.001$, 242 pairs); does not remove a shock that differentially moves one contract. \\
Mechanical rate denominator & Remove the focal odds-citing article from topical exposure & Corresponding estimates are $0.276$ and $0.274$; the result is not created by counting the outcome in its denominator. \\
Same-day timing ambiguity & Strictly-later-day citation variant & Pooled $\tcite=0.239$, $p=0.001$; category pattern unchanged. \\
One-period overfit & 2024 discovery / 2025 confirmatory split & Four of seven primary category cells (Sports, Politics, Finance, Crypto) replicate under the replication rule; Tech's sign is not concordant across the split. \\
Matcher false positives & Stratified blinded precision audit (303/303, independently double-coded) & Odds-citation precision $0.987$ and attribution precision $0.976$ (0.977/0.962 under the second coder); timing-dependent recall error remains possible. \\
Selective reporting & Full 756-cell grid and BH correction of the seven-cell family & Main claims are tied to the primary family and the confirmatory set, with remaining cells released as robustness results. \\
\bottomrule
\end{tabular}
\caption{What the robustness design addresses, and what it does not. The table is deliberately explicit
about residual confounding rather than treating a collection of placebos as causal identification.}
\label{tab:robustness}
\end{table*}

\paragraph{Shared news and reverse ordering.} The most important alternative mechanism is a common cause. A debate, poll, injury, court ruling, or economic release can generate concentrated trading and topical coverage at the same time. Two parts of the design address this. The same-market baseline removes differences in how newsworthy each market is on average, so a market that is simply always in the news cannot create the contrast. The matched comparison then tests whether a general burst of topic news explains the rest, by pairing each dislocated market with contracts in the same topic and week that did not cross the threshold; the dislocated markets are still cited more (contrast $0.434$, $p=0.001$). What neither design can remove is an event that moves one market more strongly than its matched controls in the same week. Because that residual remains, we avoid verbs such as ``drives,'' ``causes,'' or ``induces'' when describing $\tcite$.

The ordering could also run the other way. Traders may respond to reporting that appears before the timestamp our corpus captures, or journalists and traders may both watch a source GDELT does not index. Daily horizons and imperfect publication timestamps mean we cannot say who moved first at any finer time scale. As a partial check, we examine the 630 cited events with a recorded first-citation time. 74\% of first citations come after the dislocation, and 58\% follow it by more than a day. About a quarter fall on the event's own calendar day and, given how coarse the timestamps are, could precede the move. A variant that counts only citations on strictly later calendar days isolates this fraction. Excluding the 200 same-day citations from the numerator leaves the pooled estimate positive and significant ($\tcite=0.239$, $\exp(\tcite)\approx1.27$, permutation $p=0.001$), with the category pattern unchanged (full family in the same-day exclusion paragraph below). Same-day timing therefore does not create the coupling, though it cannot rule out a shared shock arriving on later days. Still, a post-move citation spike, even one robust to several baselines, is not enough to infer that the market moved the media.

\paragraph{Event-window dependence.} The pooled statistic averages over many windows, and windows from the same market are not independent. We collapse adjacent qualifying intervals into a single episode and aggregate to a market-week panel, which keeps one market from contributing many overlapping three-day windows as if they were separate evidence. Averaging over markets and clustering standard errors by market further down-weight repeated markets. One article can still enter more than one market's windows, because the retrieval shortlist links an article to several markets. The topic-week permutation does not preserve this cross-market dependence, which would tend to make the permutation $p$-values slightly too small. We therefore treat the permutation and matched-control $p$-values as screening evidence and rest the stronger claim on the 2025 confirmatory set.

\paragraph{Measurement error.} The matcher audit shows high precision, but it samples items the machine flagged as positive; it was independently double-coded, yet it bounds precision, not recall. Missed citations are harder to quantify. Misses spread evenly over time reduce power; misses or false alarms correlated with events can bias the contrast. The confidence-stratum analyses and the hand-built retrieval set make it unlikely that the headline rides on obviously weak matches. We include the annotated candidate set and instructions in the released artifact and report citation detection separately from market attribution.

The impact model has a different selection problem. Only 43\% of fitted tokens yield detectable positive impact. The non-detectable set mixes deep markets, sparse markets, nonlinear impact, and noisy estimates. Calling those tokens infinitely expensive would overstate how resistant the platform is; calling them free would understate it. All leverage analyses therefore use the same detectable population and compare bins within it.

\paragraph{Multiplicity and temporal generalization.} The full grid contains many reasonable specifications. The seven-cell primary family and BH correction keep the narrative from being picked out of all 756 cells, and the 2025 confirmatory set tests whether the pattern survives into a later period. Four replications are evidence of partial stability, not a guarantee the pattern holds elsewhere. Platform design, user composition, and newsroom practice changed between 2024 and 2025 and kept changing after the formal media partnerships. We therefore do not project the measured rates beyond the observation window.

\paragraph{Outcome form and pseudocount sensitivity.} The primary rate outcome applies a log transform with a pseudocount $c=0.5$, and we verify that the headline is neither an artifact of $c$ nor of the rate construction itself (Table~\ref{tab:outcomeform}). Varying $c$ over $\{0.1,0.25,0.5,1.0\}$ moves the pooled log-rate-ratio only between $+0.270$ and $+0.324$ (a $+31\%$ to $+38\%$ lift), with permutation $p=0.001$ throughout. Two outcomes that avoid the log-plus-pseudocount transformation give the same qualitative result. A denominator-free, self-controlled binary indicator of any citation is higher in dislocation weeks ($p=0.001$), and a Poisson count model with candidate exposure as a $\log$ offset (category fixed effects, market-clustered standard errors) gives an incidence-rate ratio of $2.48$ ($p=0.003$); a negative-binomial fit that relaxes the equal-dispersion assumption gives $2.55$ ($p=0.003$). The 33\% estimate is therefore not created by the log-plus-pseudocount transformation.

\begin{table*}[t]
\centering
\small
\begin{tabular}{lrr}
\toprule
Outcome / specification & Estimate & Perm./Wald $p$ \\
\midrule
Rate log-rate-ratio, $c=0.5$ (primary) & $+0.283\ (\times1.33)$ & $0.001$ \\
\quad $c=0.1$ & $+0.324\ (\times1.38)$ & $0.001$ \\
\quad $c=0.25$ & $+0.300\ (\times1.35)$ & $0.001$ \\
\quad $c=1.0$ & $+0.270\ (\times1.31)$ & $0.001$ \\
Binary incidence, self-controlled mean diff. & $+0.006$ & $0.001$ \\
Poisson count IRR (offset, cat.\ FE, clustered) & $2.48$ & $0.003$ \\
Negative-binomial count IRR & $2.55$ & $0.003$ \\
\bottomrule
\end{tabular}
\caption{Outcome-form and pseudocount robustness for the pooled PM five-point, three-day headline. Rate
rows are market-averaged log-rate-ratios (with the multiplicative lift $\exp(\tcite)$); the binary row is a
self-controlled mean difference in citation incidence; count rows are incidence-rate ratios,
$\exp(\hat\beta_{\mathrm{dislocation}})$. All are estimated on the same market-week panel.}
\label{tab:outcomeform}
\end{table*}

\paragraph{Denominator circularity.} The topical-article denominator can mechanically include the focal odds citation itself. We rerun the primary cell after removing that article from the denominator. The estimates are $0.276$ and $0.274$, and the headline is therefore not created by dividing by a denominator that contains the outcome.

\paragraph{Same-day exclusion.} Citations with date-resolution timestamps that fall on the same calendar day as a dislocation peak could precede the move. A variant drops all 200 such same-day citations (of 918 market-linked citations) from the numerator while keeping the exposure denominator canonical. The pooled estimate remains positive and significant ($\tcite=0.239$, $\exp(\tcite)\approx1.27$, permutation $p=0.001$), and the primary family keeps its pattern (Sports $+0.713$, Politics $+0.312$, Finance $+0.306$, Crypto $+0.065$, all $p=0.001$ with BH $q=0.001$; Culture $+0.008$ and Weather $+0.113$ not significant; Tech $-0.077$). This check was run after registration and is labeled exploratory.

\paragraph{Pre-specified equivalence test for the liquidity coefficient.} The design pre-specified a TOST equivalence test at a smallest effect size of interest of $|\beta|<0.1$ for any null claim about the liquidity coefficient in the citation-selection model. The realized coefficient is positive and significant ($\beta=0.301$), and the TOST does not find it equivalent to zero ($p=0.991$); no null claim about liquidity is made anywhere in the paper. We report the test for completeness of the registered analysis set.

\paragraph{Thresholds and horizons.} The full table includes 5- and 10-point dislocations; 1-, 3-, and 7-day horizons; count, binary, and rate outcomes; and raw and purged news controls. Main-text conclusions are restricted to patterns that do not depend on one convenient cell. Exact cell-level estimates should be read from the released grid rather than inferred from the selected figures.

\paragraph{Partnerships and editorial routines.} Two venue--outlet partnerships begin late in the 2025 window, and the later arrangements, including Polymarket's, fall after it. A partnership can raise citation regardless of how markets move, by placing odds directly inside newsroom tools. That is one mechanism through which a platform signal becomes infrastructure, not a nuisance outside the theory. The current sample is too small for a credible partner-level causal comparison. We code partnership timing and outlet status for descriptive breakdowns and treat the integration dates as a promising future quasi-experiment.

\paragraph{Partnership stratification.} Outlets are coded as partner of Polymarket, partner of a rival venue, or unaffiliated. Announcement dates are 2 December 2025 (CNN), 4 December 2025 (CNBC), 7 January 2026 (Dow Jones--Polymarket), and 17 March 2026 (Fox). Only the first two occur inside the 2024--2025 analysis period. The unaffiliated stratum is the least contract-dependent diagnostic. Among Polymarket event windows whose first citation can be classified, 613 first hits come from unaffiliated outlets and 17 from partner outlets, with none from Polymarket's own partners; these counts are too small and lopsided for a partner-level comparison.

\section{Reproducibility, data statement, and disclosure}
\label{app:release}
The artifact\footnote{\url{https://github.com/hazemibrahim97/epistemic-leverage-replication}} contains analysis scripts, fixed random seeds, the 756-cell result grid, aggregate dislocation--citation records, the pre-specified analysis-plan documents, annotation instructions, and gold-set judgments. It contains no raw copyrighted article text, no wallet-level trader labels, and no event-level tables that could be joined to public blockchain records to identify individuals. Market metadata and on-chain records remain available from their original public sources; the reused panel is cited and attributed under its CC-BY license.

\end{document}